\documentstyle[12pt,epsf]{article}
\newcommand{\beq}{\begin{equation}}
\newcommand{\eeq}{\end{equation}}
\newcommand{\beqa}{\begin{eqnarray}}
\newcommand{\eeqa}{\end{eqnarray}}
\newcommand{\ba}{\begin{array}}
\newcommand{\ea}{\end{array}}
\newcommand{\CR}{\nonumber \\}
\newcommand{\pa}{\partial}
\newcommand{\A}{\alpha}

\newcommand{\D}{\delta}

\newcommand{\E}{\epsilon}

\newcommand{\p}{\Phi}

\newcommand{\lm}{\lambda}

\newcommand{\la}{{\langle}}
\newcommand{\ra}{{\rangle}}
\newcommand{\half}{{1\over 2}}
\newcommand{\Tr}{{\rm Tr}}

\newcommand{\cO}{{\cal O}}
\newcommand{\cL}{{\cal L}}
\newcommand{\hF}{{\hat F}}
\newcommand{\hA}{{\hat A}}
\newcommand{\hD}{{\hat D}}
\newcommand{\T}{\theta}

\newcommand{\eq}{\begin{equation}}
\newcommand{\en}{\end{equation}}
\newcommand{\eqn}{\begin{eqnarray}}
\newcommand{\enn}{\end{eqnarray}}
\newcommand{\th}{\theta}

\newcommand{\Str}{ {\rm Str} }
\newcommand{\tr}{ {\rm tr} }

\begin{document}

\makeatletter
\def\setcaption#1{\def\@captype{#1}}
\makeatother

\begin{titlepage}
\null
\begin{flushright} 
hep-th/0006058  \\
UT-896  \\
June, 2000
\end{flushright}
\vspace{0.5cm} 
\begin{center}
{\LARGE 
The Non-Abelian Born-Infeld Action and 
Noncommutative gauge theory
\par}
\lineskip .75em
\vskip2.5cm
\normalsize
{\large Seiji Terashima}\footnote{
E-mail:\ \ seiji@hep-th.phys.s.u-tokyo.ac.jp} 
\vskip 1.5em
{\large \it  Department of Physics, Faculty of Science, University of Tokyo\\
Tokyo 113-0033, Japan}
\vskip3cm
{\bf Abstract}
\end{center} \par
In this paper we explicitly show
the equivalence between 
the non-Abelian Born-Infeld action,
which was proposed by
Tseytlin as an effective action
on several D-branes,
and its noncommutative counterpart
for slowly varying fields.
This confirms the equivalence between
the two descriptions 
of the D-branes 
using an ordinary gauge theory with a constant $B$ field
background and a noncommutative gauge theory,
claimed by Seiberg and Witten.
We also construct the general forms of 
the $2 n$-derivative terms for non-Abelian gauge fields
which are consistent with the equivalence 
in the approximation of 
neglecting $(2 n+2)$-derivative terms.

\end{titlepage}

\baselineskip=0.7cm

\section{Introduction}

\renewcommand{\theequation}{1.\arabic{equation}}\setcounter{equation}{0}

It has been known that the effective theory on D-branes in a 
background $B$ field has 
two descriptions which are an ordinary gauge theory and
a noncommutative gauge theory \cite{CoDoSc} \cite{DoHu}.
To relate these,
Seiberg and Witten proposed that
the noncommutative gauge theory is indeed equivalent to
the ordinary gauge theory
by a field redefinition, called Seiberg-Witten map \cite{SeWi}.

In the case of a D-brane, it has been known that 
the effective action on the brane is Born-Infeld action\footnote{
More precisely, an effective action on a D-brane in the approximation
becomes Dirac-Born-Infeld action \cite{Le}. 
However the part of the action, which is independent of the adjoint
scalars,
has the same form as the Born-Infeld action.
Since only this part will be used in this paper,
we will not distinguish the two actions.
}if all derivative terms are neglected \cite{FrTs}-\cite{AbCaNaYo}.\footnote{
By derivative terms, we mean terms with $n$-derivatives acting on
field strengths (not on gauge fields).}
Thus the Born-Infeld action should be consistent 
with the equivalence in this approximation.
In fact this was shown in \cite{SeWi} 
by constructing a family of actions parameterized by 
a parameter of noncommutativity, $\th$.
The family of actions contain the ordinary Born-Infeld action with 
a constant $B$ background
and its noncommutative counterpart without it
and it was shown that the family of the actions 
is $\th$-independent in the approximation.
Moreover, in \cite{Te2} \cite{Ok} it has been shown that 
the D-brane action computed in the superstring theory
is consistent with the equivalence up to two derivative terms.

The effective action of several D-branes is very important
in the recent development of the understanding of the non-perturbative
superstring theory, such as Matrix theory \cite{BaFiShSu} and 
AdS/CFT correspondence \cite{Ma}.
Although, Tseytlin proposed \cite{Ts} that
if all derivative terms are neglected,
the effective action on the branes is
a non-Abelian generalization of the Born-Infeld action
using the symmetrized trace over the Chan-Paton indices,
the effective action of the D-branes has not been understood completely.
Thus it may be important to 
establish 
the equivalence between the noncommutative and the ordinary
descriptions in this non-Abelian case
because the equivalence may provide a tool to
derive the effective action of the D-branes.

In this paper we explicitly show
the equivalence between 
the non-Abelian Born-Infeld action 
and its noncommutative counterpart
in the approximation of 
neglecting derivative terms,
using the differential equation
which (partially) defines the Seiberg-Witten map.
This is regarded as a non-trivial test of the equivalence
for the non-Abelian case.
To show this, it is important to keep
the ordering of the field strengths which are $N \times N$ matrices,
where $N$ is a number of the D-branes.
From this, the expansion with respect to 
a noncommutative parameter $\th$
is not relevant for this case.
Then, we compute a difference between an action
parameterized by $\th$ and one by $\th+\D \th$
exactly and then show that it contains at least a derivative term.
This implies that the noncommutative 
action is equivalent to the ordinary action 
in the approximation.
We also construct general forms of 
the $2 n$-derivative terms for non-Abelian gauge fields
which are consistent with the equivalence 
in the approximation of 
neglecting $(2 n+2)$-derivative terms as in the abelian case \cite{Te2}.

We note that 
it has been shown in \cite{OkTe} \cite{Ok} that 
for the case of a D-brane in the bosonic string,
we should modify the field redefinition 
by gauge-invariant but $B$-dependent correction terms
involving metric to match 
the known two-derivative terms \cite{AnTs} \cite{AnTs2}, 
thus we should modify the differential equation also.
It is reasonable to take into 
account the possibility of this type of modification, however,
such modification is not expected to change the result obtained 
in this paper
in the approximation of neglecting derivative terms
since the modification may include the derivative term as 
shown in \cite{OkTe}.
This problem will be discussed in detail in section 4.

This paper is organized as follows.
In section 2, we briefly 
review the equivalence between noncommutative 
and ordinary gauge theories shown in \cite{SeWi}.
In section 3,
we show
the equivalence between 
the non-Abelian Born-Infeld action
and its noncommutative counterpart
for slowly varying fields
using the differential equation
which relates the ordinary and noncommutative 
gauge fields.
In section 4,
we show that the ambiguity in the Seiberg-Witten map 
can be ignored to prove the equivalence.
In section 5,
we also construct 
general forms of 
the $2 n$-derivative terms for non-Abelian gauge fields
which are consistent with the equivalence 
in the approximation of 
neglecting $(2 n+2)$-derivative terms.
Finally section 6 is devoted to conclusion.

\section{Noncommutative Gauge Theory}

\renewcommand{\theequation}{2.\arabic{equation}}\setcounter{equation}{0}

In this section we review the equivalence between noncommutative 
and ordinary gauge theories discussed in \cite{SeWi}.
We consider open strings in flat space, with metric $g_{ij}$,
in the presence of a constant $B_{ij}$ and with a Dp-brane.
Here we assume that $B_{ij}$ has rank $p+1$ and $B_{ij} \neq 0$
only for $i,j=1, \ldots, p+1$.
The world-sheet action is
\beq
S=\frac{1}{4 \pi \A'} \int_{\Sigma} g_{ij} \pa_a x^i \pa^a x^j
-\frac{i}{2} \int_{\pa \Sigma} B_{ij} x^i \pa_{\tau} x^j
-i \int_{\pa \Sigma} A_i(x) \pa_{\tau} x^i,
\eeq
where $\Sigma$ is the string world-sheet, $\pa_{\tau}$ is
the tangential derivative 
along the world-sheet boundary $\pa \Sigma$ and
$A_i$ is a background gauge field.
In the case that $\Sigma$ is the upper half plane parameterized by 
$-\infty \leq \tau \leq \infty$ and $0 \leq \sigma \leq \infty$, 
the propagator evaluated at boundary points is \cite{FrTs}-\cite{AbCaNaYo}
\beq
\la x^i (\tau) x^j (\tau')  \ra = -\A' (G^{-1})^{ij} \log (\tau-\tau')^2+
\frac{i}{2} {\T}^{ij} \E (\tau-\tau'),
\eeq
where $G$ and $\T$ are 
the symmetric and antisymmetric tensors defined by
\beq
(G^{-1})^{ij} +\frac{1}{2 \pi \A'} \T^{ij} =
\left( \frac{1}{g+2 \pi \A' B} \right)^{ij}.
\eeq

In the case of $N$ D-branes, we must consider 
the Chan-Paton factors and $A_i$ and $F_{ij}$ become
$N \times N$ matrices.
From considerations of the string S-matrix,
the $B$ dependence of the effective action for fixed $G$ 
can be obtained by replacing ordinary multiplication 
in the effective action for $B=0$ by the $*$ product 
defined by the formula
\beq
\left.
f(x)*g(x)=e^{\frac{i}{2} \T^{ij}
\frac{\pa}{\pa \xi^i} \frac{\pa}{\pa \zeta^j} } f(x+\xi) g(x+\zeta )
\right|_{\xi=\zeta=0}.
\eeq
Using the point splitting reguralization, 
the effective action is 
invariant under a noncommutative gauge transformation
\beq
\hat{\D} \hA_i=\hD_i \lm,
\eeq
where covariant derivative $\hD_i$ is defined as
\beq
\hD_i E(x)= \pa_i E(x)+ i  \left( E(x) * \hA_i -\hA_i * E(x) \right).
\eeq

On the other hand, using Pauli-Villars regularization, 
$S$ is invariant under ordinary gauge transformation
\beq
\D_o A_i=\pa_i \lm.
\eeq
Therefore, the effective Lagrangian obtained in this way becomes
ordinary gauge theory.
Therefore this ordinary gauge theory 
and the corresponding noncommutative gauge theory
are equivalent under the field redefinition $\hA=\hA(A)$.
Because the two different gauge invariance should satisfy
$\hA(A)+\hat{\D}_{\hat{\lm}} \hA(A)=\hA(A+\D_{\lm} A)$,
the mapping of $A$ to $\hA$ 
is obtained as a differential equation
for $\T$,
\beqa
\D \hA_i(\T) = \D \T^{kl} \frac{\pa}{\pa \T^{kl} }\hA_i(\T)
&=& -\frac{1}{4}\D \T^{kl} 
\left[ \hA_k *(\pa_l \hA_i+\hF_{li})+(\pa_l \hA_i+\hF_{li})*\hA_k  ] 
\right. \CR 
\D \hF_{ij}(\T) = \D \T^{kl} \frac{\pa}{\pa \T^{kl} }\hF_{ij}(\T)
&=& \frac{1}{4}\D \T^{kl} [ 2 \hF_{ik}* \hF_{jl}+2 \hF_{jl}* \hF_{ik} 
\CR 
&& \hspace{.5cm} 
 - \hA_k *\left( \hD_l \hF_{ij} + \pa_l \hF_{ij} \right)
-\left( \hD_l \hF_{ij} + \pa_l \hF_{ij} \right)* \hA_k ],
\label{map}
\eeqa
where
\beq
{\hF}_{ij}=\pa_i \hA_j -\pa_j \hA_i-i \hA_i * \hA_j+i \hA_j * \hA_i.
\eeq

Furthermore, in \cite{SeWi} it has been proposed that
the effective action can be written for an arbitrary values of $\T$.
More precisely 
for given physical parameters $g_s, g_{ij}$ and $B_{ij}$ and
an auxiliary parameter $\T$,
we define $G_s, G_{ij}$ and a two form $\p_{ij}$ as
\beqa
\left( \frac{1}{G+2 \pi \A' \p} \right)^{ij} &=&
-\frac{1}{2 \pi \A'} \T^{ij} 
+\left( \frac{1}{g+2 \pi \A' B} \right)^{ij} \CR
G_s &=& g_s \left(
\det \left(-\frac{1}{2 \pi \A'} \T +\frac{1}{g+2 \pi \A' B} \right)  
\det(g+2 \pi \A' B) \right)^{-\half}.
\eeqa
Then the effective action $\hat{S}(G_s,G,\p,\T ; \, \hF)$, in which
the multiplication is the $\T$-dependent $*$ product,
is actually $\T$-independent, i.e. 
$\hat{S}(G_s,G,\p,\T ; \, \hF)=S(g_s,g,B,\T=0; \, F)$.
The effective action including $\p$ may be 
obtained using a regularization which
interpolates between Pauli-Villars and point splitting as in \cite{AnDo}.
In this paper, we simply assume
this proposal.


In the rest of this section, we consider a single D-brane.
In the approximation of 
neglecting the derivative terms, 
the effective Lagrangian is the 
Dirac-Born-Infeld Lagrangian
\beq
\cL_{DBI}=\frac{1}{g_s (2 \pi)^p (\A')^{\frac{p+1}{2}}}
\sqrt{\det(g+2 \pi \A' (B+F))},
\eeq
where $F_{ij}=\pa_i A_j -\pa_j A_i$.
Here $g_s$ is the closed string coupling and 
the normalization of the Lagrangian is same 
as the one taken in \cite{SeWi}.
Therefore the equivalent noncommutative gauge theory
in the approximation has the following Lagrangian
\beq
\hat{\cL}_{DBI}=\frac{1}{G_s (2 \pi)^p (\A')^{\frac{p+1}{2}}}
\sqrt{\det(G+2 \pi \A' \hF)}.
\label{Ln}
\eeq
Note that all the multiplication entering the 
r.h.s of (\ref{Ln}) can be regarded as 
the ordinary multiplication except those in the definition of $\hF$
because of the approximation.
From the requirement $\cL_{DBI}=\hat{\cL}_{DBI}$ for $F=0$,
the overall normalization $G_s$ should be fixed 
as $G_s=g_s \sqrt{\det(G)/ \det(g+2 \pi \A' B)}$.

In the approximation of neglecting the derivative of $F$,
the equation 
\beq
\D \cL_{\p}= 
\left. \D \T^{kl} \frac{\pa \cL_{\p} }{\pa \T^{kl}} 
\right|_{g_s,g,B,A_i \, fixed}
= {\rm total} \,\, {\rm derivative}
+{\cO} (\pa^2),
\label{phiLd}
\eeq
should hold, where 
\beq
\D=\D \T^{kl} \frac{\pa }{\pa \T^{kl}}.
\eeq
Here $\cL_{\p}$ is the Lagrangian defined as
\beq
\cL_{\p}=\frac{1}{G_s (2 \pi)^p (\A')^{\frac{p+1}{2}}}
\sqrt{\det(G+2 \pi \A' (\hF+\p))},
\label{phiL}
\eeq
where the multiplication is the $*$ product except in the definition of $\hF$.
Below for simplicity we set $2 \pi \A'=1$.
The variation of $G_s, G$ and $\p$ are 
\beqa
\D G_s &=& \frac{1}{2} G_s \Tr ( \p \D \T ), \CR
\D G &=& G \D \T \p + \p \D \T G, \CR
\D \p &=& \p \D \T \p + G \D \T G,
\eeqa
and the variation of $\hF$ is
\beqa
\D \hF_{ij} &=& -( \hF \D \T \hF)_{ij} -\hA_k \D \T^{kl} 
\frac{1}{2} (\pa_l+\hD_l) \hF_{ij} +\cO (\pa^4) \CR
&=& -( \hF \D \T \hF)_{ij} -\hA_k \D \T^{kl} 
(\pa_l -\half \T^{mn} \pa_n \hA_l \pa_m ) \hF_{ij} +\cO (\pa^4).
\eeqa

Following \cite{SeWi}, we get
\beqa
\!\!\!\!\!\!\!\!\! &&  \!\!\!\!\! 
\D \left( \frac{1}{G_s} \det (G+\hF+\p)^{\half} \right) \CR
\!\!\!\!\!\!\!\!\! &&\!\!  =-\half \frac{1}{G_s} \det (G+\hF+\p)^{\half} 
\left( \Tr (\hF \D \T ) +\!\! \left( \frac{1}{G+\hF+\p} \right)_{ji} 
\!\! \hA_k \D \T^{kl} \half (\pa_l +\hD_l) \hF_{ij} \right) ,
\eeqa
where the multiplication is the ordinary one 
except in $\hF$ and $\hD_l$.
Now using
\beq
\half (\pa_l+\hD_l) \hA_k-
\half (\pa_k+\hD_k) \hA_l = 
\hD_l \hA_k -\pa_k \hA_l=\hF_{lk},
\label{daf}
\eeq
we see that
\beqa
&& \!\!\!\!\! \D \T^{kl} (\pa_l+\hD_l) 
\left( \hA_k \det (G+\hF+\p)^{\half} \right) \CR
&& \hspace{0cm} =\D \T^{kl} \det (G+\hF+\p)^{\half} 
\left( \hF_{lk} +\!\!\half \!\!\left( \frac{1}{G+\hF+\p} \right)_{ji} 
\!\! \hA_k (\pa_l+\hD_l) \hF_{ij} \right)\!\! +\!\! \cO (\pa^4),
\eeqa
is a total derivative.
Thus we obtain the desired result
\beq
\D \left( \frac{1}{G_s} \det (G+\hF+\p)^{\half} \right)=
{\rm total} \,\, {\rm derivative} +\cO (\pa^4).
\label{comp}
\eeq

\section{The Equivalence between non-Abelian Born-Infeld Actions}

\renewcommand{\theequation}{3.\arabic{equation}}\setcounter{equation}{0}

In this section, 
we consider the non-Abelian ($N \times N$ matrix valued) 
gauge field $A_i$ 
on the $N$ D-branes.
In this case,
we should keep the ordering of $F_{ij}$ in the action because of the 
non-Abelian nature of $F_{ij}$ even for $B=0$.
However,
we will see that a noncommutative extension 
of the non-Abelian Born-Infeld action
satisfies the equivalence in the approximation
of neglecting derivative terms.


Let us consider 
the non-Abelian Born-Infeld action proposed by Tseytlin \cite{Ts}
\beq
\cL_{NBI}=c_p \,
\Str_{\{ F_{ij} \}} \,
\sqrt{\det(g_{ij}+(B+F)_{ij})},
\eeq
where the determinant is computed with respect to the 
worldvolume indices ${i,j}$ only and 
${\Str}_{\{ F_{ij} \}}$ means to symmetrize with respect to $F_{ij}$
and to take trace over the Chan-Paton indices,
\beq
{\Str}_{\{ F_{ij} \}} (F_{i_1 j_1} \cdots F_{i_n j_n} ) \equiv
\tr \left( {\rm Sym}_{\{F_{ij}\}} (F_{i_1 j_1} \cdots F_{i_n j_n} ) 
\right),
\eeq
where
\beq
{\rm Sym}_{\{F_{ij}\}} (F_{i_1 j_1} \cdots F_{i_n j_n} ) \equiv
\frac{1}{n !} (F_{i_1 j_1} \cdots F_{i_n j_n} 
+ {\rm all \, permutations} ),
\eeq
and ${c_p}^{-1}=g_s (2 \pi)^p (\A')^{\frac{p+1}{2}}=
g_s (2 \pi)^{\frac{p-1}{2}}$.
Here $\det$ and square root should be taken 
as if $F_{ij}$ is not the $N \times N$ matrix but 
a number and $\tr$ is the trace over the $N \times N$ 
Chan-Paton indices. 
In other words, we should forget the ordering of $F_{ij}$.
Note that the ambiguity of the ordering is fixed by the 
symmetrization.
An explicit form of $\cL_{NBI}$ is 
\beqa
\cL_{NBI} &=& c_p \, \tr \, \left[
1+\frac{1}{4} F_{ij} F_{ij} 
-\frac{1}{96} \left(
8 F_{ij} F_{kj} F_{il} F_{kl} + 4 F_{ij} F_{kj} F_{kl} F_{il} 
\right. \right. \CR
&& \left. \left. \hspace{4.5cm}
-2 F_{ij} F_{ij} F_{kl} F_{kl} -F_{ij} F_{kl} F_{ij} F_{kl} 
\right) +\cdots \right],
\eeqa
where we set $g_{ij}=\D_{ij}$ for notational simplicity.

In \cite{Ts}, it was argued that 
this non-Abelian Born-Infeld action becomes
the effective action on the $N$ D-branes 
when we neglect the covariant derivative terms $D_k F_{ij}$.
Here we should treat the commutator term $[F_{kl},F_{ij}]$
as a derivative term
since $[D_k, D_l] F_{ij}= -i [F_{kl},F_{ij}]$.

A noncommutative extension 
of the non-Abelian Born-Infeld action 
$\cL_{NBI*}$ is the Lagrangian defined as
\beq
\cL_{NBI*} = \Str_{\{ \hat{F}_{ij} \}} \, \left(  \cL_{\p}  \right) ,
\label{clp}
\eeq
where
\beq
\cL_{\p}=\frac{1}{G_s} 
\left[
{{\det}_*  (G+\hat{F}+\p) }  \right]^\half_*
=  \frac{ {\det} G^\half} {G_s} \,
\left(  {{\det}_*  (1+G^{-1} (\hat{F}+\p) ) }  \right)^\half_*,
\label{phiL1}
\eeq
The multiplication in (\ref{phiL1}) is the $*$ product
as indicated by $\left(  \det_* (\cdots) \right)^\half_*$.
We will not explicitly indicate the $*$ product below
since we will always use the $*$ product.
Note that the $*$ product is reduced to the ordinary product
for constant fields.

In this noncommutative case,
we will regard 
$\hD_k \hF_{ij} $ as 
a derivative term for an arbitrary $\th$.
In order to see that 
this is natural,
we will prove a claim 
that 
if $D_k F_{ij} =0$ for any $i,j,k$ at $\th=0$,
then $ \hD_k \hF_{ij}  =0$ for any $\th$ and any $i,j,k$.
The derivation of $\hD \hF$ with respect to $\th$
can be computed as
\beqa
\D \left( \hD_k \hF_{ij} \right) \!\!\!\! &=& \!\! \frac{1}{4} \D \th^{pq}
\left( 2 \hD_k \{ \hF_{ip}, \hF_{jq} \} +2 \{ \hD_q \hF_{ij}, \hF_{pk} \} 
- \{ \hA_p, (\hD_q+\pa_q) \hD_k \hF_{ij} \}
\right).
\label{dDF}
\eeqa
The r.h.s. of (\ref{dDF}) vanishes if $\hD_k \hF_{ij}  =0$,
which implies that the above claim is true.

Let us compute the variation with respect to $\cL_{NBI*}$ with respect to 
$\th$
\beq
\D \cL_{NBI*}= 
\left. \D \T^{kl} \frac{\pa \cL_{NBI*} }{\pa \T^{kl}} 
\right|_{g_s,g,B,A_i \, fixed} = \Delta_{\p} +\Delta_{*},
\label{phiLd1}
\eeq
where $g_s,g_{ij},B_{ij}$ and $A_i$ are fixed.
Here  the term $\Delta_{\p}$ 
includes the contributions from
$ \D G_s$, $\D G_{ij}$, $\D \p_{ij}$ and $\D \hat{F}_{ij}$ and
the term $\Delta_{*}$ includes the contributions from
the variation of $\th$ in the $*$ product.

Using the property of the $\Str$, 
\beq
\D \, \left( \Str_{\{ \hat{F}_{ij}\} } 
\left( \cL_{\p} (\hat{F}_{ij}) \right) \right)= 
  \Str_{\{\hat{F}_{ij}, \D \hat{F}_{ij} \}} 
\left( \sum_{k,l} \D \hat{F}_{kl} 
\frac{\pa}{\pa \hat{F}_{kl} } \cL_{\p} (\hat{F}_{ij}) \right)
+\cdots,
\eeq
where the ellipsis denotes 
the contributions from the variation of $\th$
in the $*$ product corresponding to $\Delta_*$,
we can find
\beqa
\Delta_{\p} &=& \half
\Str_{\{ \hat{F}_{ij}, \D \hat{F}_{ij}  \}} \, \left[
\cL_{\p}
\, \Tr \left( \D \th \p +\frac{1}{1+G^{-1} (\hat{F}+\p) } 
\left( \D G^{-1} (\hat{F}+\p) + G^{-1} (\D \hat{F}+\D \p) \right)
\right)
\right] \CR
&=& \half
\Str_{\{ \hat{F}_{ij}, \D \hat{F}_{ij}  \}} \, \left[
\cL_{\p} 
\Tr \!\! \left( \frac{1}{1+G^{-1} (\hat{F}+\p) } 
\left( \D \th (G+\p) \!-\! G^{-1} \p \D \th \hat{F}+G^{-1} \D \hat{F}  \right)
\right)
\right].
\eeqa
Here $\Tr$ is trace over worldvolume indices and $\D \hat{F}$ is evaluated 
after taking the symmetrized trace.
Note that we can insert 
the $N \times N$ identity matrix $\{1\}_{ij}
=\{ \frac{1}{1+G^{-1} (\hat{F}+\p)} 
(1+G^{-1} (\hat{F}+\p)) \}_{ij}$ into the $\Str
(\Tr (\cdots))$ without changing the result.
Substituting $\D \th (G+\p ) = \D \th G(1+G^{-1} 
(\hat{F}+\p))-\D \th \hat{F}$ and
$\Tr (\D \th G)=0$ into $\Delta_{\p}$,
we see 
\beqa
\Delta_{\p} &=& \half  
\Str_{\{ \hat{F}_{ij}, \D \hat{F}_{ij}  \}} \, \left[
\cL_{\p} \, \Tr \left( \frac{1}{1+G^{-1} (\hat{F}+\p) } 
\left( -\D \th \hat{F} -G^{-1} \p \D \th \hat{F}+G^{-1} \D \hat{F} \right)
\right)
\right] \CR
&=& \half  
\Str_{\{ \hF_{ij}, \D \hF_{ij}  \}} \, \left[
\cL_{\p} \, \Tr \left( \frac{1}{1+G^{-1} (\hF+\p) } 
G^{-1} \left(\hF \D \th \hF + \D \hF \right)-\D \th \hF
\right)
\right].
\label{dphi}
\eeqa

As in the abelian case \cite{SeWi}, 
we consider to add a total derivative term
\beqa
\Delta_{t.d.} &\equiv  & 
\half  \D \T^{kl} \tr \left( (\pa_l+\hD_l) \, 
{\rm Sym}_{\{ \hA_{i}, \hF_{ij}  \}} \,
\left( \hA_k  \cL_{\p}     \right)  \right) \CR
&=& 
\half  
\Str_{\{ \hA_{i}, \hF_{ij},  (\pa_l+D_l) \hF_{ji} \}} 
\Biggl[
\cL_{\p} \Biggl( \half \left( \frac{1}{1+G^{-1} (\hF+\p) } G^{-1} 
\right)^{ij}
\D \th_{kl} \hA_k (\pa_l+\hD_l) \hF_{ji} 
\CR
&&  \hspace{4cm}
+\Tr (\D \th \hF) 
\Biggr)
\Biggr],
\eeqa
to $\D \cL_{NBI*}$.
Here we have used (\ref{daf}) which is valid for the non-Abelian fields.
Let us define differential operators 
${\D'}$ and $\tilde{\D}$ as
\beqa
&&  {\D'} \hF_{ij} \equiv 
-\frac{1}{2} \left[  (\hF \D \th  \hF)_{ij}- (\hF \D \th  \hF)_{ji}
\right], \CR
&&  {\D'} G_s={\D'} G_{ij}
={\D'} \p_{ij}={\D'} \th =0,
\eeqa
and
\beq
\tilde{\D} E = \frac{1}{4} \D \T^{kl} 
\left\{ \hA_k \, , \,  (\pa_l+\hD_l) E \right\},
\eeq
where $E$ is an arbitrary function of $\hA_i$
and $\{  \, , \, \}$ is the anticommutator.
The operator ${\D'}$, which is
supposed to satisfy Leibniz rule, 
should act only a function of $\hF$ which contains
neither $\hA_i$ nor $\hD_l$ explicitly.
Note that 
$({\D'}-\tilde{\D})$ is not equivalent to $ \D$
though $({\D'}-\tilde{\D} ) \hF_{ij}= \D \hF_{ij}$.
Then substituting $\D \hF_{ij}=({\D'}-\tilde{\D} ) \hF_{ij}$ into
$\D \hF$ in (\ref{dphi}),
we can see
\beqa
\Delta_{\p} +\Delta_{t.d.} &=  & 
\half  
\Str_{\{ \hF, \D \hF , \hA, (\pa+\hD)\hF) \}} \, \Bigg[
\cL_{\p} \, 
\left( \frac{1}{1+G^{-1} (\hF+\p) } G^{-1} 
\right)^{ij} \CR
& & \hspace{1cm} \times
\left( \D \hF_{ij} +\half ( \hF \D \th \hF)_{ij} + \half ( \hF \D \th \hF)_{ji} + 
\frac{1}{4} \D \T^{kl} 
\left\{ \hA_k \, , \,  (\pa_l+\hD_l) \hF_{ji} \right\}  
\right)
\Bigg] \CR
&=  &  ({\D'}-\tilde{\D}) \left(
\Str_{\{  \hF_{ij}  \}} \,
\left[
\cL_{\p} \right] \right) 
- \left(
\Str_{\{  \hA_i, \hF_{ij}, (\pa_l+\hD_l) \hF_{ji}  \}} \,
\left[
({\D'}-\tilde{\D}) \cL_{\p} \right] \right) \CR
&=  &
{\D'} \left(
\Str_{\{  \hF_{ij}  \}} \,
\left[ \cL_{\p} \right] \right) 
- \left( \Str_{\{  \hF_{ij} \}} \,
\left[ {\D'} \cL_{\p} \right] \right) \CR
&  &
-\tilde{\D} \left(
\Str_{\{  \hF_{ij}  \}} \,
\left[ \cL_{\p} \right] \right) 
+ \left( \Str_{\{  \hA_i,\hF_{ij}, (\pa_l+\hD_l) \hF_{ji} \}} \,
\left[ \tilde{\D} \cL_{\p} \right] \right).
\label{d1d2}
\eeqa
In the above equation, 
${\D'} \left(
\Str_{\{  \hF_{ij}  \}} \, \left[
\cL_{\p} \right] \right) - \left(
\Str_{\{  \hF_{ij} \}} \, \left[
{\D'} \cL_{\p} \right] \right)$ 
would vanish if we neglect the ordering of $\hF$.
Thus this can be expressed as a sum of the polynomials of $\hF$ 
which contain at least a commutator of $\hF$'s
and is considered to be derivative terms in the sense
of \cite{Ts}.
However,
the last line $-\tilde{\D} \left(
\Str_{\{  \hF_{ij}  \}} \,
\left[ \cL_{\p} \right] \right) 
+ \left( \Str_{\{  \hA_i,\hF_{ij}, (\pa_l+\hD_l) \hF_{ji} \}} \,
\left[ \tilde{\D} \cL_{\p} \right] \right)$ 
is not considered to be derivative terms
because it contains
$\pa_l \hF$ and $\hA_i$.
To proceed further,
we expand this in $\hF$ and concentrate on a term 
\beq
L=-\tilde{\D} \left(
\Str_{\{  \hF_{ij}  \}} \,
\left[ \hF_1 \hF_2 \cdots \hF_n \right] \right) 
+ \left( \Str_{\{  \hA_i,\hF_{ij}, (\pa_l+\hD_l) \hF_{ji} \}} \,
\left[ \tilde{\D} (\hF_1 \hF_2 \cdots \hF_n) \right] \right),
\label{fff}
\eeq
where $\hF_l=\hF_{i_l j_l}$.
Moving $\hA_i$ in the second term of (\ref{fff}) to 
the head of the term and
extracting the terms which have the form $\hF_1 \hF_2 \cdots \hF_n$
concerning the ordering of $\hF$ from (\ref{fff}),
we find
\beqa
L \!\! &=& \!  -\frac{1}{4} \D \th^{kl} 
\tr \left[ 
\sum_{p=1}^n 
\hF_1 \cdots \hF_{p-1} 
(\hA_k (\hD_l+\pa_l) \hF_p+((\hD_l+\pa_l)\hF_p) \hA_k) 
\, \hF_{p+1} \cdots \hF_n  \right] \CR
&& \hspace{.0cm} +\frac{1}{2} \D \th^{kl} 
\tr \left[  \hA_k
\sum_{p=1}^n 
\hF_1 \cdots \hF_{p-1} 
((\hD_l+\pa_l)\hF_p) \hF_{p+1} \cdots \hF_n \right]
+{\rm total \, derivative} \CR
&=& \!\!  \frac{i}{4} \D \th^{kl} 
\tr \left[ 
\sum_{p=2}^n 
((\hD_k-\pa_k)(\hF_1 \cdots \hF_{p-1})) \, ((\hD_l+\pa_l) \hF_p) 
\, \hF_{p+1} \cdots \hF_n  \right.\CR
&& \left. \hspace{1.7cm} -\sum_{p=1}^{n-1} 
\hF_1 \cdots \hF_{p-1} ((\hD_l+\pa_l)\hF_p) \,
 (\hD_k-\pa_k ) ( \hF_{p+1} \cdots \hF_n)  \right] 
+{\rm total \, derivative} \CR
&=& \!\!  \frac{i}{2} \D \th^{kl} 
\tr 
\sum_{p=2}^n \left[ \,\,  
( \hD_k (\hF_1 \cdots \hF_{p-1})) \, (\hD_l \hF_p) 
\, \hF_{p+1} \cdots \hF_n \right. \CR
&& \hspace{2.3cm} \left. - (\pa_k (\hF_1 \cdots \hF_{p-1})) \, 
(\pa_l \hF_p) 
\, \hF_{p+1} \cdots \hF_n \,\,
\right] \,\, +{\rm total \, derivative}, 
\label{Ld}
\eeqa
where we have used $[\hA_k, \hF]=i (\hD_k-\pa_k) \hF$.
Note that the variation $\D$ of the $*$ product can be read from
\beq
\D (X_1 \cdots X_n) = \frac{i}{2} \D \th^{kl} 
\sum_{p=2}^n ( \pa_k(X_1 \cdots X^{p-1}))\,
 (\pa_l X^p) X^{p+1} \cdots X^{n},
\label{ds}
\eeq
where $\D X_i=0$.
Then the second term in (\ref{Ld}) is canceled 
by the contribution from $\Delta_{*}$.

Therefore 
from (\ref{d1d2}), (\ref{Ld}) and (\ref{ds}),
we finally obtain that
\beqa
\D \cL_{NBI*} &=& 
\Delta_{\p} +\Delta_{*}- \Delta_{t.d.}+{\rm total \, derivative} \CR
&=&  {\D'} \left(
\Str_{\{  \hF_{ij}  \}} \,
\left[ \cL_{\p} \right] \right) 
- \left( \Str_{\{  \hF_{ij} \}} \,
\left[ {\D'} \cL_{\p} \right] \right) \CR
&& \hspace{1cm}  
+{\D}_\hD \left( \Str_{\{  \hF_{ij}  \}} \,
\left[ \cL_{\p} \right] \right) 
-{\rm total \, derivative},
\label{phiLd2}
\eeqa
where the linear operator 
${\D}_\hD$ is defined as
\beq
{\D}_\hD  (\hF_1 \cdots \hF_n) = \frac{i}{2} \D \th^{kl} 
\sum_{p=2}^n ( \hD_k (\hF_1 \cdots \hF_{p-1}))\, 
(\hD_l \hF_p) \hF_{p+1} \cdots \hF_{n}.
\label{d3s}
\eeq
Since (\ref{phiLd2}) does not contain $\hA_i$ or $\pa_k \hF_{ij}$,
$\D \cL_{NBI*}$ is derivative terms in the sense of
\cite{Ts} plus total derivative terms.
Therefore we conclude that 
the non-Abelian Born-Infeld action satisfies
the equivalence in the approximation of neglecting derivative terms.

We note that (\ref{phiLd2}) 
does not contains $\cO (\hD^2 \hF^2)$ terms since
$\D \th^{kl} (\hD_k \hF_{ij}) (\hD_l \hF_{ij})=0$ and 
${\D'} \left(
\Str_{\{  \hF_{ij}  \}} \,
\left[ \hF_{ij} \hF_{ij} \right] \right) 
- \left( \Str_{\{  \hF_{ij} \}} \,
\left[ {\D'} (\hF_{ij} \hF_{ij} ) \right] \right)=0$.
However, (\ref{phiLd2}) may contain $\cO (\hD^2 \hF^{2n})$ terms,
where $n >1$, then
the non-Abelian Born-Infeld action itself does not satisfies 
the equivalence 
and some derivative corrections or/and modification of the Seiberg-Witten
map (\ref{map}) should be needed.
It is an interesting problem to find such corrections
and to compare those with the result obtained in \cite{HaTa}.

\section{Ambiguity in the Seiberg-Witten Map}

\renewcommand{\theequation}{4.\arabic{equation}}\setcounter{equation}{0}

Here we will shortly discuss the ambiguity in the Seiberg-Witten map 
and its effect on the above proof of the equivalence.
When we regard (\ref{map}) as the partial differential equation,
it is not integrable.
Thus the solution of it depends on the path in the 
$\theta$ space
although the path-dependence is absorbed by the 
gauge transformation and the field redefinition at fixed $\theta$,
as explicitly shown in \cite{AsKi}.
In fact, by reconsidering the derivation of (\ref{map}),
we can see that (\ref{map}) will not be imposed for all $\th$.
The equation (\ref{map}) should be imposed at each $\th$ 
only modulo the gauge transformation and
the field redefinition at fixed $\th$.
Then, strictly speaking, (\ref{map}) should
not be regarded as the partial differential equation
and there are ambiguities in the solution of 
the equation (\ref{map}) modulo these.

However, the ambiguities 
are not relevant for the proof of the 
equivalence.
The reason is the following.
At first order of $\D \th$
the ambiguity arising from the gauge transformation discussed in 
\cite{AsKi} is 
\beqa
\D \hF_{ij}(\T) 
&=& \frac{1}{4}\D \T^{kl} \Big( 2 \hF_{ik}* \hF_{jl}+2 \hF_{jl}* \hF_{ik} 
- \hA_k *\left( \hD_l \hF_{ij} + \pa_l \hF_{ij} \right)
-\left( \hD_l \hF_{ij} + \pa_l \hF_{ij} \right)* \hA_k 
\CR 
&& \hspace{.5cm} 
 - i \Big[ \hF_{ij}, \A \hF_{kl} +\beta [\hA_k,\hA_l \, ] \, \Big] \Big).
\eeqa
We can easily see that the last term 
does not contribute the $\D \cL_{NBI}$ since
it has the form of the gauge transformation.
The ambiguity arising from the field redefinition 
should have the form 
$\D \hA_i \sim \D \T^{kl} H_{ikl}(G,\T,\p,\hF, \hD \hF, \hD \hD \hF,\cdots)$ 
because of the gauge invariance.
Note that the number of $\hD$ in $H_{ikl}$ is odd.
Thus the contributions from this term to $\D \hF$ is 
$\D \hF_{ij}\ \sim \D \T^{kl} (\hD_i H_{jkl} -\hD_j H_{ikl} )$ 
in the first order of $\D \th$.
Therefore the corrections to $\D \cL_{NBI}$ from this term 
are derivative terms and
we conclude that we can ignore the ambiguities 
to prove the equivalence in the approximation.

In order to prove the equivalence including the higher derivative terms,
we should take into account th ambiguity.
From the fact that 
the $\th$ is not appear explicitly in (\ref{phiLd2}) 
and $\D ({\rm derivative \, corrections})$,
we can find the possible form of the Seiberg-Witten map is 
\beqa
\D \hF_{ij}(\T) 
&=& \frac{1}{4}\D \T^{kl} \Big( 2 \hF_{ik}* \hF_{jl}+2 \hF_{jl}* \hF_{ik} 
- \hA_k *\left( \hD_l \hF_{ij} + \pa_l \hF_{ij} \right)
-\left( \hD_l \hF_{ij} + \pa_l \hF_{ij} \right)* \hA_k 
\CR 
&& \hspace{.5cm} 
 - (\hD_i H_{jkl} -\hD_j H_{ikl}) \Big),
\eeqa
where 
\beq
H_{ikl}= H_{ikl}(G^{-1},\hF+\p, \hD \hF, \hD \hD \hF,\cdots).
\label{hikl}
\eeq
Note that the (\ref{hikl}) depends on $\alpha'$
although it is not explicitly indicated.

\section{Derivative Corrections}

\renewcommand{\theequation}{5.\arabic{equation}}\setcounter{equation}{0}

In this section, we consider the derivative terms which 
are consistent with the equivalence.
The tree level effective action of 
the D-branes in the superstring theory
is expected to be the non-Abelian Born-Infeld action with
an appropriate linear combination of these derivative terms
because the effective action should satisfy the equivalence.
Thus it is desired to 
find the general forms of the derivative corrections 
which satisfy the equivalence.
As a first step to find them,
we will construct the $2 m$-derivative terms which satisfy the equivalence
in the approximation of neglecting $(2m+2)$-derivative terms.

The general forms of the derivative corrections 
which satisfy the equivalence in the approximation 
were obtained in \cite{Te2}
for the abelian case.
Below we will generalize the result obtained in \cite{Te2}
to the non-Abelian case.
To do this,
we define 
\beqa
{(h_S)}^{ij} &=& \left( \frac{1}{G+\hF+\p} \right)^{ij}_{\rm sym}=
\half \left( \frac{1}{G+\hF+\p} \right)^{ij}+
\half \left( \frac{1}{G-\hF-\p} \right)^{ij} \CR
&=& \left( \frac{1}{G+\hF+\p} G \frac{1}{G-\hF-\p} \right)^{ij},
\eeqa
and denote an arbitrary $(0,3m )$ tensor
of a form
\beq
\left\{ (\hD \hF) \cdots (\hD \hF)
 \right\}_{p_1 p_2 \cdots p_{3m}},
\eeq
by $ J_{p_1 p_2 \cdots p_{3m}} $.
For example, we can take $J_{p_1 p_2 \cdots p_{6}}=
\hD_{p_1} \hF_{p_2 p_3} \hD_{p_4} \hF_{p_5 p_6}$ for $m=2$.
This tensor will be used only in $\Str$ or ${\rm Sym}$,
then the ordering of $\hD \hF$ in it will fixed.

Now we consider the $m$-derivative terms 
\beq
\cL_{m} = \Str_{\{ \hF_{ij} , \hD_k \hF_{ij}\}} 
\, \left(  \cL_{\p}  \, L_{m} \right) ,
\label{clm}
\eeq
where 
\beq
L_m = 
(h_S)^{p_1 p_2} (h_S)^{p_3 p_4} \cdots (h_S)^{p_{3m-1} p_{3m}}
\,\, J_{p_1 p_2 \cdots p_{3m}},
\eeq
is a $m$-derivative terms and 
$\cL_{\p}$ is
the $\th$-dependent non-Abelian Born-Infeld Lagrangian 
defined in (\ref{phiL1}).
We separate the variation of $\cL_{m}$ with respect to $\th$ 
to three parts such that
\beq
\D \cL_{m}= \Delta_{\p}^{m}+\Delta_{L}^{m}+\Delta_{*}^{m},
\eeq
where 
$\Delta_{\p}^{m}$ and $\Delta_{L}^{m}$ are the contributions 
from $ \D G_s, \D G_{ij}, \D \p_{ij}, \D \hF_{ij}$ and $\D (\hD_k \hF_{ij})$ 
in $\cL_{\p}$ and in $L_m$,
respectively,
and $\Delta_{*}^{m}$ comes from the variation of $\th$ 
in the $*$ product except in $\hF$ and $\hD \hF$.

Next we will compute $(\D+\tilde{\D}) (\hD_k  \hF)$.
We can show that
\beqa
[\D, \hD_k] E &=& -\frac{i}{4} \th^{pq} \left(
\left\{ [E,\hA_p], (\pa_q \hA_k+\hF_{qk})  \right\}+
\{ \hA_p, [E,\pa_q \hA_k+\hF_{qk}]\} 
\right) \CR
&& -\frac{1}{2} \D \th^{pq} (\pa_p E \pa_q \hA_k -\pa_p \hA_k \pa_q E),
\eeqa
where $\hD_k E= \pa_k E +i [E,\hA_k]$, and
\beq
[\tilde{\D}, \hD_k] E =\frac{1}{4} \th^{pq} \left(
-\left\{ \hD_k \hA_p, (\pa_q+\hD_q) E\right\}+
i \left\{ \hA_p, [ E, \hF_{qk}+\pa_q \hA_k ] \right\}
\right).
\eeq
From these, after some computations we find a simple result
\beq
[\D+\tilde{\D}, \hD_k] E = 
-\frac{1}{2} \D \th^{pq} \{ \hF_{kp}, \hD_q E \}.
\eeq
Then using 
\beq
(\D+\tilde{\D}) \hF_{ij}= 
-\frac{1}{2} \D \th^{pq} \{ \hF_{ip}, \hF_{qj} \} ,
\eeq
we obtain 
\beqa
(\D+\tilde{\D}) (\hD_k \hF_{ij}) &= &
-\frac{1}{2} \D \th^{pq} 
\left( \{ \hD_k \hF_{ip}, \hF_{qj} \}+\{ \hF_{ip}, \hD_k \hF_{qj} \}
+\{ \hF_{kp}, \hD_q \hF_{ij} \}
\right) \CR
&= & -\frac{1}{2} \left( \hD_k \{ \hF\D  \th, \hF\}_{ij} +
\{ (\hF \D \th)_{k}^{\,\, q} , \hD_q \hF_{ij} \} \right).
\eeqa
As in the abelian case \cite{Te2}, 
we also show that
\beq
(\D+\tilde{\D}) (h_S)^{ij} = 
\left( h_S (\hF \D \th) +(\D \th \hF ) h_S \right)^{ij}
+\D_{*} (h_S)^{ij}+\cdots,
\eeq
and then
\beq
(\D+\tilde{\D}) L_m = 
\D_{*} L_m+\cdots,
\eeq
where 
$\D_{*} (h_S)^{ij}$ and $\D_{*} L_m$ are contributions 
coming from the variation of 
$\th$ in the $*$ product and
the ellipsis denotes terms involving a commutator $\hF$,
which are regarded as the $(m+2)$-derivative terms.

According to the discussion in the previous section,
we finally find 
\beq
\D \cL_m+\Delta_{t.d}^{m}
=  \Delta_{m+2} + {\rm total \, derivative},
\label{diffL2}
\eeq
where 
$\Delta_{m+2}$ is a $(m+2)$-derivative term
and
\beq
\Delta_{t.d}^{m} \equiv 
\half \D \T^{kl} \tr \left( (\pa_l+\hD_l) \,
{\rm Sym}_{\{ \hA,\hF,\hD \hF \}} \, 
\left(  \hA_k \cL_{\p}  \, L_{m} \right) \right),
\eeq
is a total derivative term.
Therefore the $m$-derivative correction (\ref{clm})
satisfies the equivalence in the approximation of
neglecting $(m+2)$-derivative terms.

In \cite{Te2}, it was shown that 
a type of derivative corrections containing $h_{A}$,
which is defined as
\beqa
{(h_A)}^{ij} &=& \left( \frac{1}{G+\hF+\p} \right)^{ij}_{\rm anti sym}=
\half \left( \frac{1}{G+\hF+\p} \right)^{ij}-
\half \left( \frac{1}{G-\hF-\p} \right)^{ij} \CR
&=& -\left( \frac{1}{G+\hF+\p} (\hF+\p) \frac{1}{G-\hF-\p} \right)^{ij},
\eeqa
also satisfies the equivalence.
As in the above discussion on the derivative correction containing $h_S$,
we can easily shown that the generalization of 
this type of derivative corrections to the non-Abelian gauge fields
also satisfies the equivalence.

\section{Conclusion}

\renewcommand{\theequation}{6.\arabic{equation}}\setcounter{equation}{0}

We have shown that 
the non-Abelian Born-Infeld action 
is equivalent to its noncommutative counterpart
in the approximation of neglecting derivative terms
not expanding the action with respect to 
the noncommutative parameter $\th$.
We have also constructed the general forms of 
the $2 n$-derivative terms for the non-Abelian gauge fields
which are consistent with the equivalence 
in the approximation of 
neglecting $(2 n+2)$-derivative terms.
It may capture some general structures of the 
effective action of the D-branes.

It is interesting to generalize
the results obtained in this paper to
construct the action which satisfies the equivalence
without the approximation of neglecting derivative terms,
which may has applications, especially, for a relation between 
the nonlinear instanton \cite{SeWi} \cite{Te} \cite{MaMiMoSt}
and the noncommutative instanton \cite{NeSh}.
Since we should treat the non-Abelian gauge fields,
there is the ordering problem
even for the ordinary gauge fields, which has not been solved yet.
Thus the constraints using the equivalence 
are expected to be important for 
determination of the effective action on the several D-branes.
If we success to construct such an action,
we would solve the ordering problem also.

To supplement this approach to obtaining the effective action of D-branes,
it would be important to 
consider the supersymmetric extension of the action.
The superfields in noncommutative geometry has been discussed in 
\cite{Te3}-\cite{ChZa} and the supersymmetric 
non-Abelian Dirac-Born-Infeld action in noncommutative geometry
was discussed in \cite{GrPaSc}.
Then it might be interesting to consider 
supersymmetric noncommutative gauge theories and
their equivalence relations.

The simplified Seiberg-Witten map \cite{CoSc} \cite{Is}
may be also useful to
construct the action consistent with the equivalence 
as in \cite{Co}.
Although the simplified Seiberg-Witten map
is different from the Seiberg-Witten map 
in the higher order of $\th$, 
the derivative corrections obtained in 
\cite{Te2} using the Seiberg-Witten map
coincide with those obtained in \cite{Co}.
In order to proceed this method further,
it is important 
to study the relation between the two maps.

\vskip6mm\noindent
{\bf Acknowledgements}

\vskip2mm
I would like to thank T. Kawano and Y. Okawa
for useful discussions.
This work was supported in part by JSPS Research Fellowships for Young 
Scientists. \\

\noindent
{\bf Note added}: 

While preparing this article for publication,
we received the preprint
\cite{Co2} which 
discussed the general structure of the 
non-Abelian Born-Infeld action from 
the equivalence between ordinary and noncommutative gauge theories
using an algebraic method.
In particular,
in two dimension the non-Abelian Born-Infeld action
was recovered from the equivalence
and its lowest derivative correction was found.
On the other hand, the result obtained in this paper 
does not depend on the dimension of the D-branes.

\newpage


\end{document}